\documentclass[prl,twocolumn,floatfix]{revtex4}
\usepackage[dvips]{graphics}

\begin{document}
\title{Granular clustering in a hydrodynamic simulation}
\author{Scott A. Hill}
\author{Gene F. Mazenko}
\affiliation{James Franck Institute and Department of Physics, University of Chicago, Chicago, Illinois 60637.}
\date{\today}
\begin{abstract}
We present a numerical simulation of a granular material using
hydrodynamic equations.  We show that, in the absence of external
forces, such a system phase-separates into high density and low
density regions.  We show that this separation is dependent on the
inelasticity of collisions, and comment on the mechanism for this
clustering behavior.  Our results are compatible with the granular
clustering seen in experiments and molecular dynamic simulations of
inelastic hard disks.
\end{abstract}
\maketitle

One of the key differences between a granular material and a regular
fluid is that the grains of the former lose energy with each
collision, while the molecules of the latter do not.  Even when the
inelasticity of the collisions is small, it can give rise to dramatic
effects, including the \textit{Maxwell Demon} effect\cite{eggers} and,
the topic of this paper, the phenomenon of granular clustering.
Experiments\cite{kudrolli,olafsen} and molecular dynamics
simulations\cite{goldhirsch} alike show that granular gases in the
absence of gravity do not become homogeneous with time, but instead
form dense clusters of stationary particles surrounded by a lower
density region of more energetic particles.  From a particulate point
of view, one can explain these clusters by noting that when a particle
enters a region of slightly higher density, it has more collisions,
loses more energy, and so is less able to leave that region, thus
increasing the local density and making it more likely for the next
particle to be captured.  We, however, are interested in describing
this behavior using hydrodynamics.  There is considerable
work\cite{kinetic} deriving granular hydrodynamics from kinetic
theory, focusing on analytical treatments of the long-wavelength
behavior of the system.  Goldhirsch and Zannetti\cite{goldhirsch}, for
instance, describe clustering as the result of a hydrodynamic
instability: a region of slightly higher density has more collisions,
and thus has a lower temperature.  A lower temperature means a lower
pressure, which attracts more mass from the surrounding
higher-pressure region.  Their long-wavelength stability analysis
shows that in a system of hydrodynamic equations similar to
Eq.~\ref{eq-eqns}, higher-density regions do indeed have lower
pressure, fueling the instability.  Our approach here is different.
In this paper, we discuss whether a coarse-grained description, in
terms of local particle, momentum, and energy densities, can be used
to treat characteristic behaviors of granular materials as a
self-contained dynamical system.  The spirit is similar to the use of
density functional theory in the treatment of freezing\cite{freezing}.
We present the results of numerical simulations for such a description
in the case of clustering, and show that when inelasticity is present,
one indeed sees the system phase-separate into regions of high and low
density.

We begin with a number density field $\rho$, a flow velocity field
$\mathbf{u}$, and a ``temperature''\cite{temperature} field $T$.
These are related by a standard set of hydrodynamic equations for
granular materials\cite{haff}:

\begin{eqnarray}
\label{eq-eqns}
{\partial\rho\over\partial t}&=&-\nabla_i(\rho u_i)\cr
{\partial(\rho u_i)\over\partial t}&=&-\nabla_i P
-\nabla_j(\rho u_iu_j)+\nabla_j(\eta_{ijkl}\nabla_ku_l)\cr
{\partial T\over\partial t}&=&-\nabla_i(u_iT)+{1\over\rho}\nabla_i(\kappa \nabla_i T)\cr
&&\hskip0.5in+\frac{1}{\rho}\eta_{ijkl}(\nabla_iu_j)(\nabla_ku_l)-\gamma T.
\end{eqnarray}
where repeated indices are summed over, and where $P$ is the pressure,
$\kappa$ is the bare thermal conductivity,
 and $\eta_{ijkl}=\eta(\delta_{ik}\delta_{jl}+\delta_{il}\delta_{kj}+\delta_{ij}\delta_{kl})$
is the isotropic bare viscosity tensor.  These equations bear
much in common with those for normal fluids\cite{kim}.  The most
important addition is that of the inelastic term, $-\gamma T$.  The
parameter $\gamma$ is a measure of the inelasticity of the collisions,
and is proportional to $(1-e^2)$, where $e$ is the coefficient of
restitution.  Using kinetic theory results\cite{haff}, the transport
coefficients are chosen to depend on temperature and density:
\begin{eqnarray}
\kappa=\kappa_0 T^{1/2}\cr
\eta=\eta_0 \rho T^{1/2}\cr
\gamma=\gamma_0 T^{1/2}.
\end{eqnarray}

Typically, work in granular hydrodynamics is done in low-density
regimes, where grains may be treated as point particles interacting
via collisions.  When simulating aggregation, however, one must take
excluded volume into account.  We do this by introducing a barrier in
the pressure $P(\rho)$ at some maximum (close-packed) density
$\rho_0$.  This is in addition to the usual hydrodynamic pressure
$\rho T$.  We choose in particular the simple quadratic form
\begin{equation}
P=\rho T+U(\rho^2-\rho_0^2)\theta(\rho-\rho_0),
\end{equation}
where $U$ is a positive parameter, $\theta(x)$ is the unit step
function, and $\rho_0$ is the close-packed density.
We introduced this method in an earlier paper\cite{hill}, and it is a
simple way to model the incompressibility of the system at high
densities\cite{gollub}.  

We evaluate our equations in two dimensions using a finite-difference
Runge-Kutta method, on a $100\times100$ square lattice with periodic
boundary conditions.  (See footnote \cite{method} for more details.)
The lattice spacing is chosen to be large enough so that each site
contains a number of grains, and we can consider the density to be a
continuous variable.  We start with random initial conditions
$\rho=0.1+0.001r_1(z,x)$, $u_z=r_2(z,x)$, $u_x=r_3(z,x)$, and
$T=1+0.1r_4(z,x)$, where $r_i(z,x)$ are random numbers chosen between
$-1$ and $1$.  The model's other parameters for the data presented
here are $\eta_0=50$, $\kappa_0=1$, $U=4\times 10^4$, $\rho_0=0.2$,
and $\gamma_0=50$ or $0$.  All values are in arbitrary units, with a
time step of $\Delta t=10^{-3}$.

\begin{figure}[!ht]
\begin{center}\includegraphics{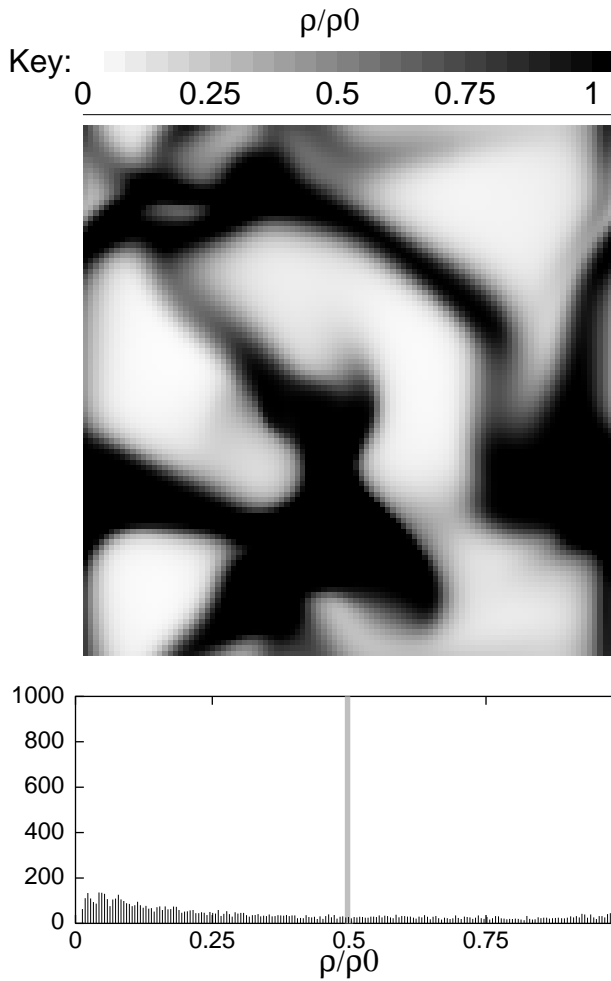}\end{center}
\caption{\label{fig-gam} The density distribution at $t=800$ when
$\gamma_0=50$, clearly showing phase separation.  The top picture is a
spatial distribution of the density; darker shades correspond to
higher densities.  The lower graph is a histogram of the density, with
bin size $\Delta(\rho/\rho_0)=0.005$.  The initial distribution is
shown by the light gray bar in the center.}
\end{figure}

\begin{figure}[!htp]\begin{center}\includegraphics{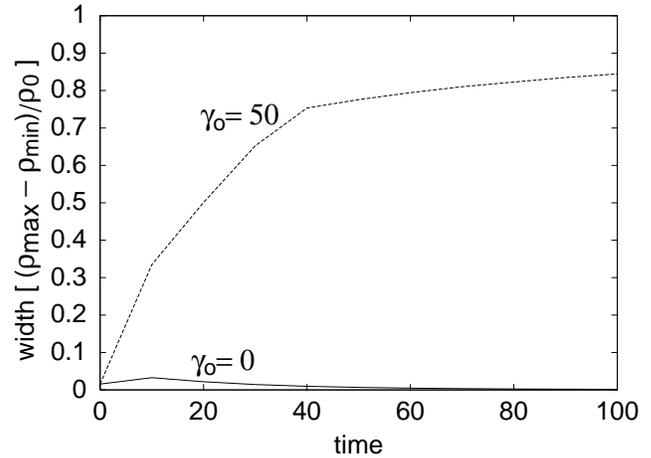}\end{center}
\caption{\label{fig-cmp}The width of the density distributions,
simply defined as the difference between the maximum density and the
minimum density, for $\gamma_0=0$ and $\gamma_0=50$.}
\end{figure}

Figure~\ref{fig-gam} shows the density distribution for the system
after it has evolved to a time $t=800$. One can clearly see that, from
a narrow initial distribution centered at $\rho/\rho_0=0.5$, the
system has separated into regions of high and low density, with
relatively sharp borders.  Calculating the structure factor
$S(\mathbf{r})=\left\langle\delta\rho(\mathbf{r'})\delta\rho(\mathbf{r'}+\mathbf{r})\right\rangle_{\mathbf{r'}}$
shows that the density is correlated out to 8 or 9 lattice spacings.
To show that this clustering behavior is due to the inelastic term in
Eq.~\ref{eq-eqns}, we duplicate the run with $\gamma_0=0$.
Figure~\ref{fig-cmp} shows the width of the density distribution as a
function of time for the inelastic and elastic cases.  Without
inelasticity, the system quickly becomes homogeneous.

This clustering behavior of the system appears to be robust.  The
parameter values chosen for this run were not optimized, and a variety
of alternate choices result in the same qualitative effect.  The time
over which clustering occurs does depend on different parameters, and in
cases where the viscosity is too large, the density distribution does
not form the peaks at its extremes, but instead remains broad and
relatively flat.  This also occurs when one removes the pressure
barrier at $\rho_0$.  In both cases, however, there is distinct and
permanent phase separation.

\begin{figure}[!ht]\begin{center}\includegraphics{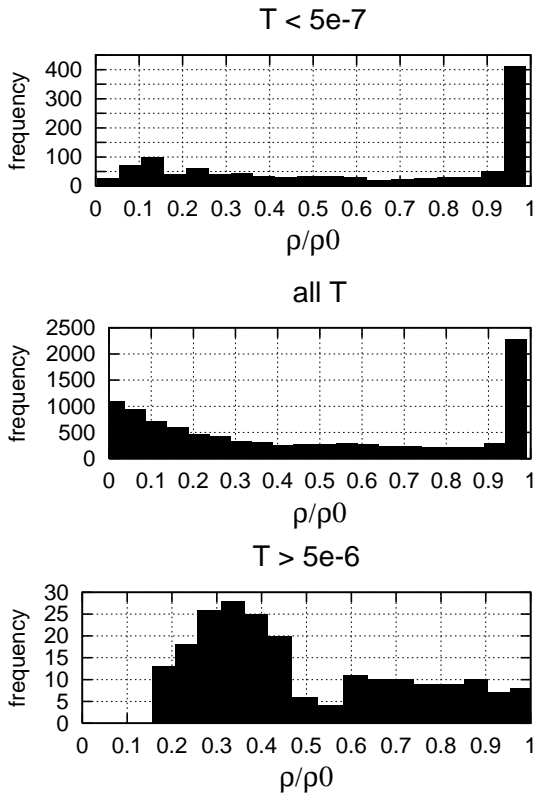}\end{center}
\caption{\label{fig-corr} Density distributions for the $\gamma_0=50$
case at $t=800$.  The middle graph is the distribution for the entire
system.  The upper graph is the distribution only for those sites on
the lattice where the temperature is less than $5\times 10^{-7}$.
(The average temperature at this time is $1.3\times 10^{-6}$.)
Similarly, the lower graph is the distribution for those sites with
temperature greater than $5\times 10^{-6}$.}
\end{figure}

\begin{figure}[!ht]\begin{center}\includegraphics{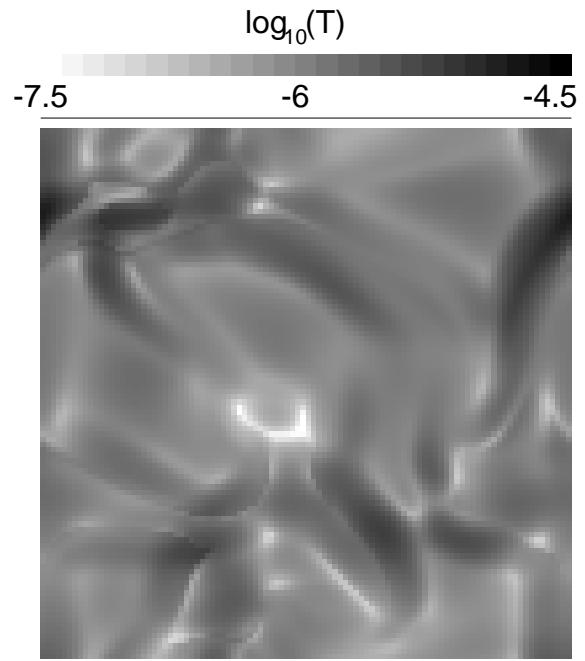}\end{center}
\caption{\label{fig-temp}The spatial distribution of the log of the
temperature at $t=800$.  Just as before, darker shades
correspond to higher temperatures.  Compare this figure with
Fig.~\ref{fig-gam} to look for correlations between density and temperature.}
\end{figure}

It is clear that our simulations reproduce clustering.  As mentioned in
the introduction, it is believed \cite{goldhirsch} that clustering
occurs because of an anticorrelation between density and temperature:
regions of high density have more collisions, and thus more
dissipation, and thus less total energy.  Fig.~\ref{fig-corr} shows
the density distributions corresponding to the highest and the lowest
temperatures.  Notice that the high-temperature distribution is
missing the high-density peak present in the other two graphs.  On the
other hand, in the low-temperature distribution, the high-density peak
is relatively larger than in the general case.  Both points suggest
that high-density regions tend to be at low temperatures, which agrees
with the common wisdom.  However, the anticorrelation is not as
clear-cut as one might expect: there is no correlation between low
densities and high temperatures, for instance, and even the high
density sites have a broad range of temperatures which only tend
toward the low side.  Fig.~\ref{fig-temp} is a density plot of the
temperature.  When one compares this with Fig.~\ref{fig-gam}, there
appears to be more structure to the system than the simple explanation
would suggest.

\begin{figure}[!ht]
\begin{center}
\includegraphics{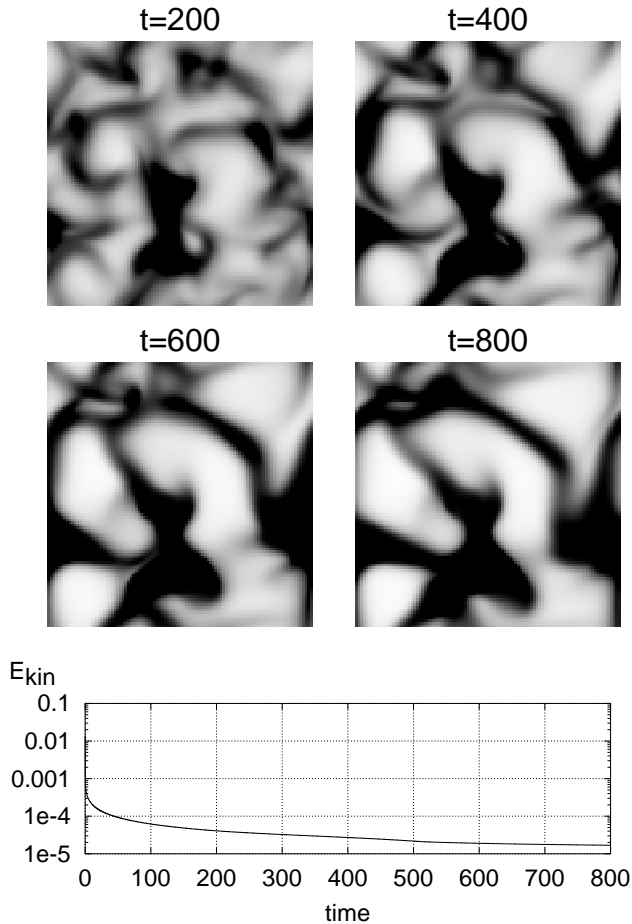}
\end{center}
\caption{\label{fig-moving}At the top, images of the spatial density
distribution at four times, using the same key as in
Fig.~\ref{fig-gam}.  Note that the clusters continue to move around for
long periods of time.  Below these, a plot showing that the average
kinetic energy ($E_{kin}=\langle \rho(u_z^2+u_x^2)\rangle$) is
decaying at a slower-than-exponential rate.}
\end{figure}

One interesting point about the system is that it takes a long time to
come to a stop.  Fig.~\ref{fig-moving} shows the distribution of
density at four different times.  Also shown is the average kinetic
energy of the system over time; it is decaying at a
slower-than-exponential rate.  This occurs because the viscosity,
being proportional to the temperature, is decaying to zero, and is
thus very small by the end of this run.  It is difficult to say
whether this feature corresponds to the results of experiments and
other simulations, because the nature of freezing in gravity-free
granular systems is generally subtle.  In event-driven molecular
dynamic simulations, one finds the phenomenon of inelastic
collapse\cite{inelastic}, where particles on the verge of freezing in
a cluster suffer an infinite number of collisions in a finite amount
of time.  In experiment\cite{olafsen}, there is the opposite effect:
one must vibrate real two-dimensional systems of particles to create
clustering, as the surface friction will otherwise freeze the system
before any clustering can occur.  If desired, one could bring our model
to a halt with the introduction of a surface friction term, or by
replacing the current temperature dependence of the viscosity with
\begin{equation}
\eta\propto \eta_0\rho(T+T_{\min})^{1/2}.
\end{equation}

In summary, we have recreated in hydrodynamic simulation the clustering
behavior in granular materials which was predicted by hydrodynamic
theory and observed in kinetic simulations.  We show that this
behavior is directly dependent on the inelasticity parameter of our
equations.  This supports the notion that granular materials can be,
to some extent at least, described by means of coarse-grained
variables in a nonlinear hydrodynamical setting.  However, we also
suggest that long-wavelength hydrodynamics alone may not be sufficient
to fully describe granular behavior, such as the actual mechanism for
granular clustering.

We thank Professor Todd DuPont for his assistance in constructing our
simulations, and Professors Heinrich Jaeger, Sidney Nagel, and Thomas
Witten for helpful conversations.  This work was supported by the
Materials Research Science and Engineering Center through Grant
No. NSF DMR 9808595.


\begin{thebibliography}{99}
\bibitem{temperature}The temperature $T$ here is not thermal, but
actually a measure of the local energy density ${\cal E}$ that is not
accounted for by the local flow velocity: $\rho T\equiv {\cal
E}-{1\over2}\rho u^2$.  Apart from this equation, it is not necessary
to assume that this quantity $T$ has any properties normally
associated with temperature.
\bibitem{method}Some details about our methods:
(1) Our Runge-Kutta method includes an adaptive time step; it,
however, is rarely activated since we use a default time step that is
suitably small.
(2) We offset the velocities from the other fields on the lattice
such that the densities and temperatures may be considered to lie on
the faces of the lattice while the horizontal (vertical) velocities
lie on the vertical (horizontal) edges.
(3) We have had great numerical difficulty with regions of very low
density: such regions have a tendency to drift below zero density or
generate large velocities in an unnatural fashion.  We use an upwind
technique to evaluate the equation $\dot\rho=-\nabla(\rho u)$, to
prevent the density from becoming negative.  In addition, we prevent
mass from flowing out of any site with a density below some minimum
value ($\rho_{\min}=0.001$).
\end{thebibliography}
\end{document}